\documentclass[aps,prb, superscriptaddress,twocolumn,amssym]{revtex4}
\usepackage{latexsym}
\usepackage{graphicx}
\usepackage{rotating}
\usepackage{longtable}
\usepackage[usenames]{color}
\usepackage[normalem]{ulem}

\input colordvi

\begin{document}

\title{Temperature dependence of surface stress across an order-disorder 
transition: $p(1\times2)$O/W(110)}

\author{N.~Stoji\'{c}}
\affiliation{Abdus Salam International Centre for Theoretical Physics, 
Strada Costiera 11, Trieste 34151, Italy}
\affiliation{ INFM-CNR Democritos, Theory @ Elettra group,  Trieste, I-34151, Italy}

\author{T.~O.~Mente\c{s}}
\affiliation{Sincrotrone Trieste S.C.p.A., Basovizza-Trieste 34012, Italy}

\author{N.~Binggeli}
\affiliation{Abdus Salam International Centre for Theoretical Physics, 
Strada Costiera 11, Trieste 34151, Italy}
\affiliation{ INFM-CNR Democritos,  Trieste, I-34151, Italy}

\author{M.~A.~Ni\~{n}o}
\affiliation{Sincrotrone Trieste S.C.p.A., Basovizza-Trieste 34012, Italy}

\author{A.~Locatelli}
\affiliation{Sincrotrone Trieste S.C.p.A., Basovizza-Trieste 34012, Italy}

\author{E.~Bauer}
\affiliation{Department of Physics, Arizona State University, Tempe, Arizona 85287-1504}

\date{\today}

\begin{abstract}
Strain relaxations of a $p(1\times2)$ ordered oxygen layer on W(110) 
are measured as a function of temperature across the disordering transition 
using low-energy electron diffraction. The measured strains approach
values of 0.027 in the $[1\bar{1}0]$ and $-0.053$ in the $[001]$ direction.  
On the basis  of the measured strain relaxations, we give quantitative 
information on temperature-dependent surface stress using the results of 
${\it ab~initio}$  calculations. From the surface formation energy for 
different strains, determined by first-principles calculations, we estimate 
that surface stress changes from $-1.1$ for the ordered phase to $-0.2$~N/m 
for the disordered one along $[1\bar{1}0]$, and from 5.1 to 3.4~N/m along 
$[001]$. Moreover, our observation that the strains scale inversely with domain 
size confirms that the strain relaxation takes place at the domain boundaries.
\end{abstract}


\maketitle

\section{Introduction}

Surface stress is a quantity important for the understanding of 
surface processes like reconstruction, interfacial mixing, segregation, 
critical film thickness and self-organization at solid 
surfaces.\cite{Hai01,Iba97,NarVan92,NieBesSte95,CamSie94} Regarding the 
latter, recent studies have demonstrated that the stress-induced patterns 
occurring at the mesoscopic scale on metal and semiconductor 
surfaces\cite{JonPelHon96,KerNieSch91,GasPlaBar03} show a strong dependence  
on temperature, \cite{MenLocAba08} suggesting that the change of surface stress 
with temperature can play a major role in determining their properties. In spite of 
the wide interest in these and other stress-driven phenomena, there are very few 
studies concerning the temperature dependence of surface stress.\cite{NDiGasMar09}

On the theoretical side, density-functional theory (DFT) is probably the most 
accurate method to study surface elastic properties; however its results apply to 
0~K.\cite{temp_0} Non-zero temperatures can be investigated using model potentials 
and atomistic simulations, but in the literature there are few examples applied 
to surface elasticity as a function of temperatures.\cite{FroMis09} More studies can be 
found on a related quantity, the surface free energy,\cite{VerWil68,Foi94,BroGil83}
which, however, does not give information on surface stresses without knowledge of 
its dependence on the surface strain. On  the experimental side, the 
crystal-bending method was used to measure stress changes due to 
adsorbed species  at different temperatures.\cite{GroErlIba95} Indirectly connected 
to surface stress, there are some measurements of temperature-dependent strain using 
x-ray diffraction. However, a truly temperature-dependent study of a given system 
is still lacking. 

Here, we aim at estimating  surface stress change as a function of temperature 
across the order-disorder transition of the $p(1\times2)$-O/W(110) structure, using 
an alternative approach which combines low-energy electron diffraction (LEED) 
and {\it ab~initio} calculations. In addition to  thermal expansion, 
the presence of the disordering is crucial in determining the surface 
stress changes, similar to the phenomenon of surface melting.\cite{FroMis09}
This stress change is also of particular importance for the recently reported
high-temperature stress-induced pattern formation of Pd on W\cite{MenLocAba08}, 
due to the  profound effect of O adsorption on the anisotropy of Pd 
stripes.\cite{note_inPreparation}

Oxygen on W(110) is a well-studied model system characterized by a series of ordered 
phases as a function of oxygen coverage.\cite{WanLuLag78,JohWilChi93} At 0.5 ML 
coverage O is ordered in a $(1\times2)$ structure,  which consists of doubly-spaced 
close-packed rows of O in the $\langle\bar{1}11\rangle$ directions. The ($1\times 2$) 
phase contains eight equivalent domains generated by translations and rotation 
(by $109.5^{\circ}$) of the O lattice on the triply coordinated  adsorption 
sites.\cite{JohWilChi93} The $p(1\times 2)$ O domains disorder at about 700~K. 
Bucholz and Lagally\cite{BucLag75} describe the disordering by shrinking 
of the ordered domains with increasing temperature. At the same time, the surface 
stress is expected to be relieved at the domain boundaries.\cite{JohWilChi93} 

In this paper, we use  LEED to detect the temperature-induced lattice 
changes in the O adlayer on W, and DFT calculations for their quantitative analysis 
in terms of surface stress. More precisely,  we estimate the surface stress by using  
surface formation energy as a function 
of strain, calculated from first principles. We evaluate the surface stress  at the 
values of strain  for which the surface layer is found to be maximally relaxed (well 
above the disordering  temperature). We focus on the behavior in a wide 
temperature range, and we show that the average change in atomic spacing of the surface 
layer for increasing temperature is a direct consequence of stress relaxation due to 
disorder on the surface. The analysis of lattice changes measured by LEED 
is done in the same way as in Ref.~\onlinecite{MenStoBin08}. Kinematic low energy 
electron diffraction, although similar to x-ray diffraction in the information sought, 
has the advantage of being inherently surface sensitive, even more so in the case of 
O/W(110) with the presence of half-order diffraction spots.

The paper is organized as follows: in section II we present the measured strain 
relaxations as a function of temperature, obtained by LEED. Section III summarizes
our ${\it ab~initio}$ calculations of surface formation energy as a function of strain. 
Combining the two, we extract estimates of surface stress at high temperature, which 
is discussed in section IV. Conclusions are given in section V.

\begin{figure}[t]
\begin{center}
\includegraphics{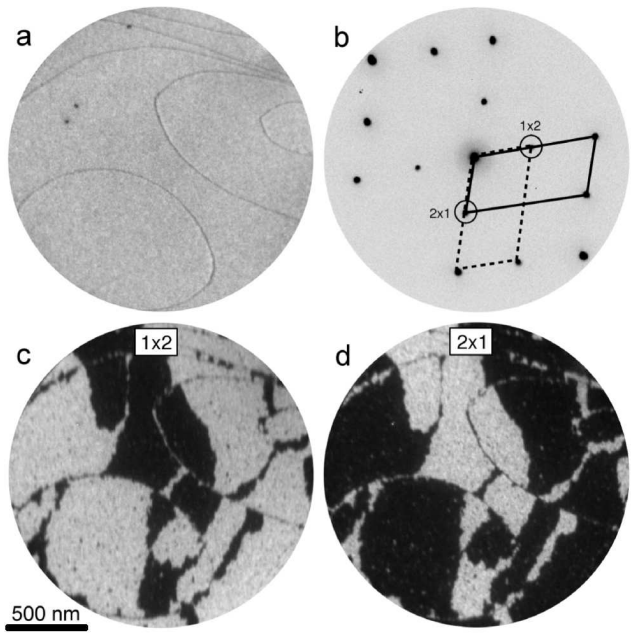}
\caption{(a) LEEM image of the partially (1x2) oxygen-covered W(110).
Image diameter is 2~microns. The curved lines correspond to W atomic steps. 
(b) LEED pattern acquired from a 2~micron region at 35 eV electron energy. 
(c) and (d) Dark field LEEM images of the same region acquired using the half-order 
spots marked in panel (b). 
The bright/dark contrast corresponds to the $1\times2$ and $2\times1$ domains. }
\label{fig:figure1x2}
\end{center}
\end{figure}

\section{Experimental results}

The LEED measurements were made with the spectroscopic photo-emission and 
low energy electron microscope SPELEEM at Elettra (Italy).\cite{LocAbaMen06} 
The instrument  is a hybrid electron/x-ray microscope, which allows chemical and
structural imaging combined with micro-spot LEED ($\mu$-LEED).\cite{note_ref,LocAbaMen06}
In the latter mode, a LEED pattern was obtained from a region of 2~$\mu$m diameter.
The transfer width of $\mu$-LEED was determined to be 10~nm from
measurements on a step-free region of the clean W(110) surface. 

Cleaning of the W(110) single crystal was carried out by annealing at 1400~K in 
$2\times10^{-6}$~mbar oxygen, followed by high temperature flashes in ultrahigh vacuum 
to remove oxygen. The surface cleanliness was confirmed by the absence of tungsten 
carbide islands and a sharp ($1\times1$) LEED pattern. Using low-energy electron 
microscopy (LEEM), regions with micron-sized terraces were chosen in order to exclude 
broadening of diffraction spots due to the presence of step bunches.

The (1$\times$2)-ordered 0.5~ML oxygen-covered surface was prepared 
by exposing the clean W(110) surface to molecular oxygen at 450~K. 0.5~ML coverage 
was assigned to the maximum intensity of the (1$\times$2) diffraction spots, which 
was obtained at an exposure of about 4.6~Langmuirs.\cite{MenStoBin08} After O exposure, 
we performed a short annealing at 1400~K under ultra-high vacuum conditions, which 
resulted in large domains with a sharp LEED pattern.

Figure~\ref{fig:figure1x2}a shows a LEEM image of the (1$\times$2)-O/W(110) surface
acquired  below the disordering of oxygen, at around 400~K. The corresponding 
$\mu$-LEED pattern is displayed in Fig.~\ref{fig:figure1x2}b. The very large  (1$\times$2) 
and (2$\times$1) domains are visible in the {\em dark field} images shown in 
Fig.~\ref{fig:figure1x2}c and \ref{fig:figure1x2}d.\cite{note_darkfield} The domain 
sizes reach micron scale with boundaries consisting of narrow regions with opposite domain
orientation between two rotational domains showing no obvious
directional anisotropy.

It should be noted that the presence or absence of translational domain boundaries 
within a region of single orientation (1$\times$2 or 2$\times$1) cannot be concluded 
from the images displayed in Fig.~\ref{fig:figure1x2}. However, the high intensity
of the half-order diffraction spots confirms the presence of extended single domains.
Nevertheless, fine lines and granularity in a single rotational domain in the high resolution
images could point to the presence of defects (translational domain boundaries, oxygen 
vacancies, etc) within an otherwise perfect oxygen lattice.

\begin{figure}[t]
\begin{center}
\includegraphics[width=7cm]{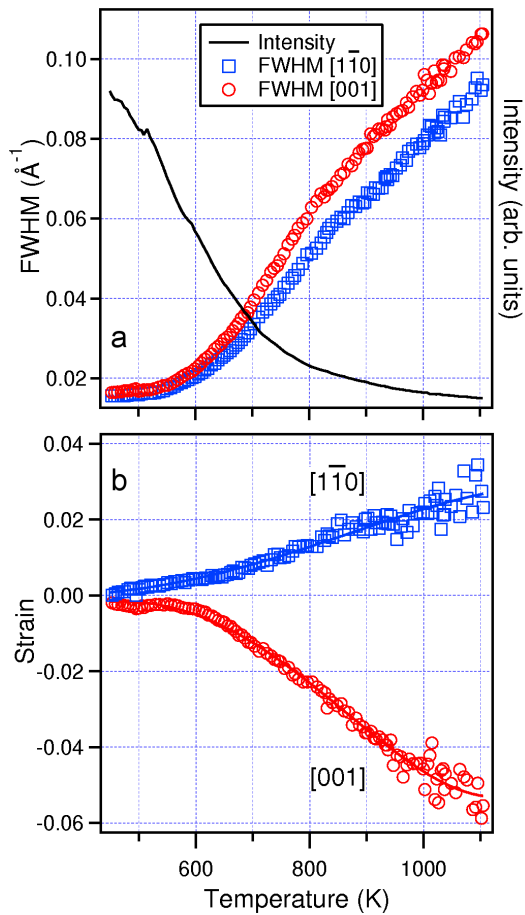}
\caption{ (a) Peak intensity (solid line) and width  of the half-order diffraction spots,
measured at electron energy of 28~eV.
(b) Measured strain, derived from the reciprocal distances of the half-order peak 
positions, as a function of temperature along two directions. Solid lines are the
least-squares fit, indicated to guide the eye.}
\label{fig:strainvsT}
\end{center}
\end{figure}

\begin{figure*}
\begin{center}
\includegraphics[width=14cm]{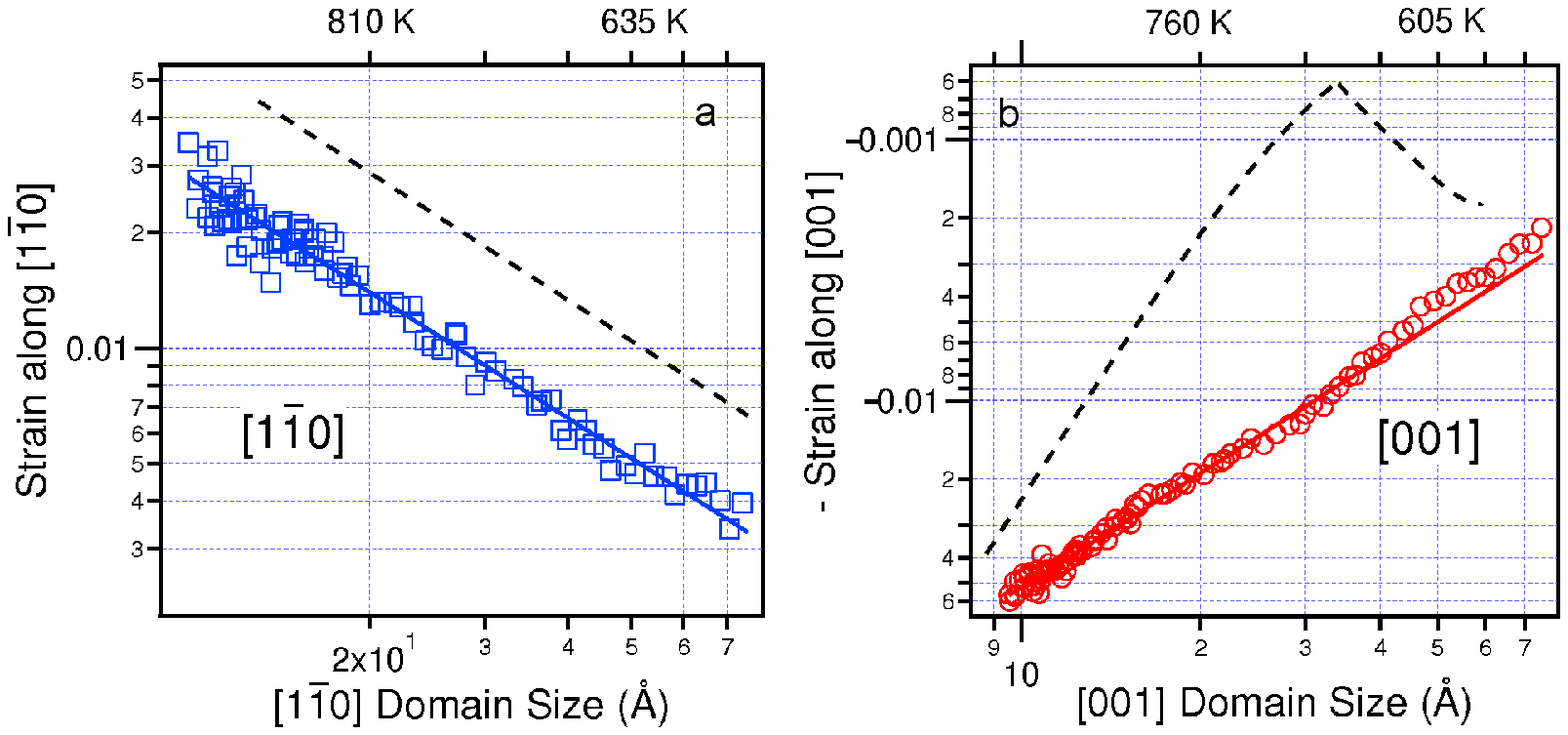}
\caption{ Changes of strain along (a) $[1\bar{1}0]$  and (b) [001] as a function of 
the average dimension of the ($1\times2$) domains. The squares and circles are the 
data extracted from the LEED measurements, while the solid lines are fits to the 
power-law functions.  Temperatures corresponding to
20, 50 and 100 \AA\ domain sizes are noted above. Strains obtained in 
Ref.~\onlinecite{MenStoBin08} as a function of oxygen coverage at fixed 
temperature are included as dashed lines for comparison.}
\label{fig:strainvsdomain}
\end{center}
\end{figure*}

This surface is used to monitor the strain 
relaxations at high temperature. The disordering of the (1$\times$2) 
order as a function of temperature can be followed via the intensity and width of the 
half-order diffraction spots displayed in Fig.~\ref{fig:strainvsT}a. The decrease 
of intensity and the increase of spot widths are documented in the early
literature~\cite{EngNieBau75,BucLag75} and can be understood as large domains 
breaking up into smaller ones with increasing temperature. This is also reflected 
in the shape of the diffraction spot profiles, which transforms from Gaussian to 
Lorentzian at about 600 K. Importantly, diffuse half-order spots can still be 
discerned up to above  1100~K. This means that the O atoms preserve the 
local (1$\times$2) order (in small clusters) even though the long-range order 
is lost at high temperature.

In addition to the intensity and width, we can also measure the half-order peak 
positions as a function of temperature. The reciprocal lattice distances allow us to 
evaluate the {\em average} distance of O atoms within the 
(1$\times$2)-ordered regions. The data analysis is carried out in the same way as 
in Ref.~\onlinecite{MenStoBin08}. The resulting strains, defined with respect 
to the underlying W lattice, in the two crystallographic 
directions are displayed in Fig.~\ref{fig:strainvsT}b. Clearly, the (1$\times$2) 
O unit cell, which respects the W periodicity at low temperatures, on average 
expands along $[1\bar{1}0]$ and shrinks along $[001]$ as the temperature increases 
and disorder sets in. At the highest temperatures the measured strains tend to
reach values of about $\epsilon_{\rm relaxed}=0.027$ in $[1\bar{1}0]$ and 
$\epsilon_{\rm relaxed}=-0.053$ in $[001]$ directions at 1100~K.

The measured values of the strain relaxation include a contribution  from the 
 thermal expansion of the W crystal. However, within the temperature range of
interest, the change in lattice spacing due to this effect is less than 
0.4~\%,\cite{coeff_ref} which is comparable to the smallest measured surface strains.

Interestingly, the strains in both directions scale as inverse power laws of the 
average domain size as shown in Fig.~\ref{fig:strainvsdomain}. Domain sizes were 
extracted from the FWHM of half-order diffraction spots as displayed in 
Fig.~\ref{fig:figure1x2}a taking proper account of the instrumental broadening.
The fit functions in the log-log plots in Fig.~\ref{fig:strainvsdomain} are described 
by $\varepsilon \propto L^{p}$, and correspond to powers of $p=-1.09$ and $-1.45$  for 
$[1\bar{1}0]$ and $[001]$ directions, respectively. This observation is in line with 
the strain relaxations taking place at the domain boundaries.
Moreover, the changes in strain as a function of O coverage at fixed temperature (450~K)
are also included in Fig.~\ref{fig:strainvsdomain} for comparison.\cite{MenStoBin08}
The important differences will be discussed in the following sections.

Note that the two main sources of error in the experiments are the limited instrument
transfer width and the changes in the angular alignment as a function of temperature.
The former makes the determination of domain sizes very difficult when the domains are
larger than the transfer width (hence the broadening of the diffraction spots is
dominated by the instrument function). However, considering that the
transfer width is about 10~nm, this should be a negligible effect within
the range displayed in Fig.~\ref{fig:strainvsdomain}.
On the other hand, variations in angular alignment may introduce a small
uncertainty in the spot positions, which might translate into slight
deviations from the power law behavior. In order to reduce the possibility
of such an artifact, we have averaged over equivalent lattice vectors
in obtaining the reported strains.

\begin{figure}[t]
\begin{center}
\includegraphics[width=8cm]{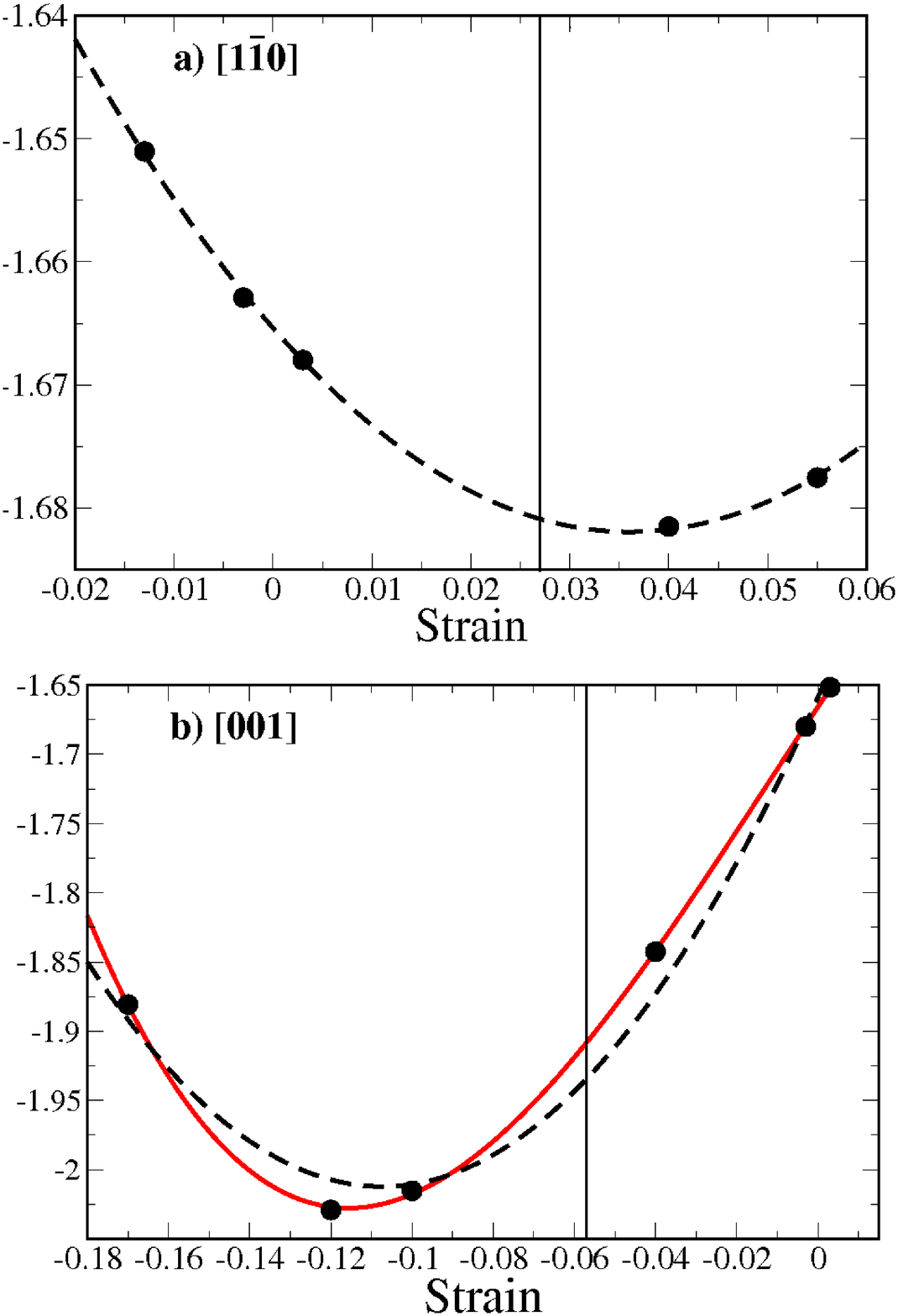}
\caption{{\it Ab initio} calculations of the surface formation energy 
multiplied by the surface unit cell area as a function of strain in a) $[1{\bar 1}0]$ 
and b) [001] directions. Black dashed curves show quadratic fit, while the solid red 
line is for a cubic fit. Full circles indicate calculated points. The vertical line 
corresponds to the measured strain, $\epsilon_{relaxed}$, of the O layer at high 
temperature.  }
\label{fig:surf_free_energy}
\end{center}
\end{figure}

\section{{\it Ab initio} calculations}

In this study, we use  {\it ab initio} calculations  to determine the surface 
formation energy, $\gamma$, for different strains. The strain derivative is used to evaluate the 
surface stress at the measured maximally-relaxed strain $\epsilon_{relaxed}$ of the O layer.

To calculate $\gamma$ we perform DFT pseudopotential calculations in a plane-wave basis 
by using the {\tt PWscf} code.\cite{GiaBarBon09} The local-density approximation (LDA) in the 
Perdew-Zunger parametrization\cite{PerZun81} is used  for the exchange and correlation 
functional. We utilize a symmetric slab with 11 layers to simulate the O/W system, with 9 
vacuum layers.  We employ Vanderbilt ultra-soft pseudopotentials,\cite{Van90} generated using 
the $2s^2 2p^4$ atomic configuration of O and the $5s^2 5p^6 5d^4 6s^2$ configuration of W. 
The core-cutoff radii  for O were: $r_{s,p}=1.6$, $r_d = 1.4$~a.u. and for W: $r_{s,p}=2.2$, 
$r_d = 2.4$~a.u. We use the equilibrium W bulk lattice constant  of 3.14~\AA. Our 
plane-wave-basis kinetic-energy cutoff is 35~Ry for the wave functions and 350~Ry for 
the charge density. We use a $26\times13\times1$ Monkhorst-Pack\cite{MonPac76} mesh for 
the  k-points sampling in the Brillouin zone of the  ($1\times2$)  surface. 

$\gamma$ is evaluated  from the expression:
\begin{equation}
\gamma = \frac{1}{2 A} (E_{\rm slab} - N_W E^{B}_W - \frac{N_O}{2} E_{O_2}),
\label{eq:gamma}
\end{equation}
where $A$ is the strain-dependent surface area, $E_{\rm slab}$ stands for the total 
energy of the whole slab, $N_W$ for the number of W atoms, $E^B_W$ for the  W bulk energy 
per atom, $N_O$ for the number of O atoms and $E_{O_2}$\cite{note_O2} is the energy of 
the $O_2$ molecule. $E_{slab}$ is evaluated for several different
homogeneous uniaxial $[1{\bar 1}0]$ and [001] strains on the slab and $E^B_W$ is 
recalculated for each strain. We note that the constant energy $E_{O_2}$ does not 
contribute to the derivative of energy and, therefore, is not relevant to the 
discussion of surface stress.

On the basis of the calculated $\gamma$ for various strains and a fitted  polynomial 
curve, the surface stress, $\tau_{i}$,\cite{note_biaxial} is obtained by using:\cite{Shu50}
\begin{equation}
\tau_i= \frac{1}{A} \frac{\partial (A \gamma)}{\partial \epsilon_i},\phantom{dsd}i=1,2
\label{eq:tau}
\end{equation}
where $\epsilon_{i}$ is the  uniaxial strain. Figure \ref{fig:surf_free_energy} 
presents $\gamma$ multiplied by the area\cite{note_area} for a few strains, as well as
polynomial fits to the calculated points for both directions. We assume that the two 
directions can be considered independently. In the $[1{\bar 1}0]$ direction, a quadratic 
function fits the data very well, while for the [001] direction, due to  
higher-order terms, a better fit is obtained by a cubic polynomial.

Calculated values of $\tau$ are presented in Table~\ref{tab:stresses}. In the first 
column are our $p(1\times2)$O/W(110) surface stress values calculated from the curves 
in Fig.~\ref{fig:surf_free_energy} at  $\epsilon=0$.  The second column contains the same 
quantities, for comparison, from Ref.~\onlinecite{MenStoBin08} (calculated with the 
expression derived by Nielsen and Martin\cite{NieMar83} for the stress tensor $\sigma$
based on the Hellmann-Feynman theorem) for both directions. Finally, in the last  column 
are our  estimates for the average surface stress in the high-temperature phase, with no 
long-range order ($\tau_{\rm dis O(1\times2)}$), as obtained from the derivative of the 
energy curve in Fig.~\ref{fig:surf_free_energy} at $\epsilon_{\rm relaxed}$.

\renewcommand{\baselinestretch}{1}
\begin{table}[ht]
\bigskip
\begin{center}
\begin{ruledtabular}
\begin{tabular}{ c c c c }
  direction & $\tau_{\rm O(1\times2)}^\gamma$
            &   \footnote{Ref.~\onlinecite{MenStoBin08}}$\tau_{\rm O(1\times2)}$
            & $\tau_{\rm dis O(1\times2)}^\gamma$ at $\epsilon_{\rm relaxed}$ \\
\hline
 $[1\bar{1}0]$         & $-1.1$       &  $-1.1$    & $-0.2$  \\
  $[001]$              & $5.1$        &  $5.4$    & $3.4$   \\
\end{tabular}
\end{ruledtabular}
\caption{Surface stresses at $\epsilon=0$  calculated from the  $\gamma$ curve 
(first column) and by a direct DFT surface stress calculation\cite{MenStoBin08}  
(second column). Last column contains  surface stresses at $\epsilon_{\rm relaxed}$, 
estimated from the $\gamma$ curve. The values are given in N/m. The positive sign 
corresponds to the tensile stress.  }
\label{tab:stresses}
\end{center}
\end{table}

\section{Discussion}

The calculated values in Table~\ref{tab:stresses} for the stresses at $\epsilon=0$ 
using the two approaches, from the $\gamma$ curve and by a direct DFT calculation of the 
surface stress, give  similar results in both directions. The differences are within 
the estimated error, 0.35 N/m,\cite{MenStoBin08} of the calculated surface stress and 
the propagated error of the fitting parameters of $\sim0.20$~N/m.  The different slab 
thicknesses in the  two calculations (15 layers for the direct stress calculation 
and 11 layers for the total energy calculation of the strained slabs) also may 
contribute to the difference. The surface stress values at $\epsilon_{\rm relaxed}$ 
(high-temperature phase) indicate a significant reduction with respect to the values 
at $\epsilon=0$ (0~K).

To better understand the physical significance of the values listed in the last 
column in Table~\ref{tab:stresses}, we need to  describe the underlying assumptions. 
Most importantly, we assume that the disordering takes place predominantly within 
the O layer and that the temperature changes in stress are mostly due to this 
disorder.
The basis for the first approximation is that W 
atoms do not participate in the ($1\times2$) order to begin 
with.\cite{VanTon75,YnzDenPal00,note_W_disturbed} Therefore, they are expected 
to be disturbed little by the disordering of the O lattice, if at all.
The second assumption is based on the small effect of the bulk thermal expansion of W.

As we noted in the introduction, with increasing temperature, the O lattice loses 
long-range order by breaking of large domains into smaller ones. It is  expected that 
the loss of order should result in a reduction of stress within the O layer with the 
stress relieved at the domain boundaries.\cite{JohWilChi93} In the limiting case of 
complete disorder (a dense gas of O atoms) one would expect negligible stress in the 
O layer, and the remaining surface stress would be mostly dominated by the underlying 
W layer. On the other hand, the presence of the half-order diffraction spots at the 
highest temperatures measured, however weak and diffuse, suggests that the O layer 
has still short-range order. Therefore, the high temperature surface stress can be 
decomposed into contributions from the unmodified W layers and from the 
{\em partially relaxed} oxygen surface.

In order to represent this situation, in our theoretical analysis, ideally, we would like to
calculate stresses for the strains within the oxygen layer equal to $\epsilon_{relax}$ and
keeping the bulk unstrained in the same slab. However, such inhomogeneous strain is not 
affordable in our supercell calculations.  Therefore, we use slabs in which both the bulk 
and the surface part are strained to $\epsilon_{relaxed}$ (with subsequent subtraction of 
the bulk contributions). To see how this theoretical treatment corresponds to the 
experiment in which we measure only the surface-layer strain  with W underneath at 
its equilibrium lattice constant (strain zero), we examine the layer-resolved stresses
as a function of strain. From Eq.~\ref{eq:gamma} and \ref{eq:tau}, the uniaxial surface 
stress can be written as:
\begin{equation}
\begin{array}{llll}
\tau_{i}(\epsilon_i)
&=\frac{1}{2} \frac{V}{A} [ \sigma_{\rm i,slab}(\epsilon_i) - N_W \sigma_{\rm i,W}^{\rm B}(\epsilon_i)]\\
&= \frac{1}{2} \frac{V}{A} [ \sigma_{\rm i,surf}(\epsilon_i) + 
    \sigma_{\rm i,subsurf}(\epsilon_i) +(N_W -1) \sigma_{i,W}^{\rm B}(\epsilon_i)\\ 
&- N_W \sigma_{\rm i,W}^{\rm B}(\epsilon_i)],
\end{array}
\label{eq:stress_layers}
\end{equation}
where $\sigma_{\rm i,slab}(\epsilon_i)=(1/V)\partial E_{\rm slab}(\epsilon_i)/\partial\epsilon_i$ 
and 
$\sigma_{\rm i,W}^{\rm B}(\epsilon_i)=(1/V)\partial E_{\rm W}^{\rm B}(\epsilon_i)/\partial\epsilon_i$ 
are the volume-averaged stresses, V is the volume of the unit cell and the indices ``surf'' 
and ``subsurf'' stand for the surface O and the subsurface W layers, respectively.  
Assuming that the subsurface layer's proximity to the surface changes its stress-strain 
relation only by addition of a constant stress (which corresponds to the stress at zero strain, 
see the Appendix), with respect to the bulk, we are left with:
\begin{equation}
\tau_i=\frac{1}{2} \frac{V}{A} \left[ \sigma_{\rm i,surf}(\epsilon_i) + 
\sigma_{\rm i, subsurf}(\epsilon=0) \right].
\label{eq:stress_const}
\end{equation}

Under the above assumption, our calculated stress at $\epsilon_{\rm relaxed}$ 
has two contributions: from the surface O layer at $\epsilon_{\rm relaxed}$ and
from the W subsurface layers at the strain zero.\cite{note_subsurf}  
We note that our assumption for Eq.~\ref{eq:stress_const} is analogous to saying 
that the elastic constants of the subsurface and bulk layers are identical. 
A heuristic interpretation of Eq.~\ref{eq:stress_layers} and \ref{eq:stress_const} 
can be found in the Appendix.

A confirmation that the elastic response of the subsurface W is close to that of bulk 
W is obtained by inspecting the relaxed positions in our calculated O/W slab. We find 
that the average interlayer spacing of the two topmost W layers below O is very close 
to its bulk value ($\Delta d_{23}=-1.1$~\%), while the clean W surface relaxes by 
$\Delta d_{12}=-3.6$~\%.\cite{MenStoBin08} This indicates that the W surface when covered 
with O becomes much more bulk-like. We note that our results for $\Delta d_{23}$ for O/W
and $\Delta d_{12}$ for W agree well with other theoretical results, 
$\Delta d_{23}=-1.3$~\% (Ref.~\onlinecite{StoPodMul09}) for the former and 
$\Delta d_{12}=-3.6$~\% (Ref.~\onlinecite{ArnHupBay97}) for the latter case. 
Experimentally,  $\Delta d_{12}=-3.1$~\% for W.\cite{ArnHupBay97}
 
In spite of the assumptions mentioned above, there are important conclusions to draw 
from the results. Interestingly, our results in Table~\ref{tab:stresses} show 
that the surface stress and  its anisotropy do not vanish even several hundred degrees 
above the disordering  transition at about 600~K.
 This is mainly due to short-range correlations within the O layer persisting at high 
temperatures. The other contribution, which may be equally significant, 
is the non-zero stress within the subsurface layers which do not participate in the 
disordering. A discrimination of the two effects needs a layer-resolved stress 
calculation, which is beyond the scope of our study.

Another point is the confirmation that stress is relieved at the domain boundaries,
which is derived from the power-law behavior in Fig.~\ref{fig:strainvsdomain}.
Although this is qualitatively similar to the relaxations at the boundaries of
small $p(1\times2)$O islands on W(110),\cite{MenStoBin08} there are differences to 
be emphasized. The strain relaxation at the boundary between two distinct phases is 
driven by the difference in surface stress, as shown in Ref.~\onlinecite{MenStoBin08}. 
However, in the case of  the fully O-covered surface, the mechanism of the stress 
relaxation is somewhat different. 

To verify that, we constructed  heavy and light domain-wall configurations
and compared their energy on the basis of broken and newly-formed bonds, using the
lateral interaction parameters for up to the third neighbors.\cite{note_boundary} 
This simple analysis yields a vacancy line (light-domain wall 
between the (1x2) and (2x1) domains) as 
the most favorable domain-wall configuration for the main crystallographic 
directions on W(110). Although the energy difference between this and some heavier 
domain walls is  smaller than 10~meV/\AA, this estimate does not include the energy 
obtained by strain relaxations, which is the largest for this type of the boundary, 
as it allows the largest  atomic relaxations. Predominant vacancy-type domain walls 
would, indeed, confirm the view that the stress is relaxed at the domain boundaries 
in this system. 
 
The differences between strain relaxation at an {\em island boundary} and a 
{\em domain boundary} are underlined in Fig.~\ref{fig:strainvsdomain}. Most importantly, 
i) along $[1{\bar 1}0]$ strain relaxation  on fully oxygen-covered surface as a 
function of domain size with increasing temperature shows a considerably reduced 
magnitude (but identical slope) compared to that of  relaxation  for small oxygen 
islands as a function of island size with increasing  coverage at fixed
temperature (dashed line in Fig.~\ref{fig:strainvsdomain}a), ii) along $[001]$ both 
magnitude and slope of the strain curves are drastically different between two data 
sets. From the first point we conclude that along $[1{\bar 1}0]$ the force driving 
the strain relaxation at the boundary is reduced for  the fully O-covered surface, 
however the relaxation mechanism remains the same. The island boundary responds 
to the stress difference between O-covered and clean tungsten, which is 
-4.7~N/m.\cite{MenStoBin08} This is, indeed, considerably larger than the stress on 
fully O-covered surface along $[1{\bar 1}0]$ (see Table~\ref{tab:stresses}) and it 
points to the idea that at the domain boundaries the relaxation is driven by the stress 
itself and not a difference of stresses. The comparison along $[001]$ gives an even stronger
support to this discussion.  In the case of an O island, the boundary almost feels no force
along $[001]$ and therefore the relaxation is a higher-order effect decaying rapidly with
increasing island size (dashed line in Fig.~\ref{fig:strainvsdomain}b). However, in the fully 
O-covered surface the strain scales with a power of domain size much closer to a 
$1/L$ power law, which is similar to the behavior along  $[1{\bar 1}0]$. This is 
in line with the idea that the domain boundary is relaxing under the influence of 
the large tensile stress along $[001]$.

We emphasize that an anisotropic thermal expansion of the oxygen layer
cannot replace surface stress as the driving force of the observed strain
relaxations. This is clearly shown in the size of the oxygen unit cell
as a function of temperature, which can be derived from the data in Fig.~\ref{fig:strainvsT}.
The oxygen unit mesh expands slightly from 400 K to 600 K, above which it
contracts almost linearly with increasing temperature. The contraction
is explained by the predominantly tensile stress within the oxygen layer.

Having shown qualitatively that the strain relaxations are due to  the stress itself and 
not the stress difference, we note that also the sign of the measured strain relaxation in 
the two directions supports this conclusion, as well as the ratio of the strain 
relaxations  (large response in the direction of larger surface stress). However, 
it is interesting to observe that the strain relaxation along  $[1\bar{1}0]$ is only 
two times smaller than along [001], and not 
five times, as could be expected from the stress ratio in the two directions.
This is explained  by different potential profiles 
in the two directions, according to which O atoms move much easier along  
$[1\bar{1}0]$, than [001].\cite{ZalKruRom01}

\section{Conclusion}

In this work, we have measured the strain relaxation as a function of temperature 
across the disordering transition of $p(1\times2)$O/W(110). The breaking up of large 
domains into smaller ones as a function of temperature can be observed by the changes 
of the intensity and width of the half-order diffraction spots. Following the 
changes of the spot distances we extracted the strain relaxation as a function of 
temperature. At T$\sim 1100$~K the measured strains reach values of 
$0.027$ along $[1\bar{1}0]$ and $-0.053$ along $[001]$. We  found that the measured 
stresses are roughly proportional to the inverse domain size. By means of 
${\it ab~initio}$  calculations, we determined the surface formation energy 
as a function of strain, from which we  estimated the surface stress values at 
the measured relaxed strains.

\section{Appendix}

\begin{figure}[t]
\begin{center}
\includegraphics[width=7cm,angle=0]{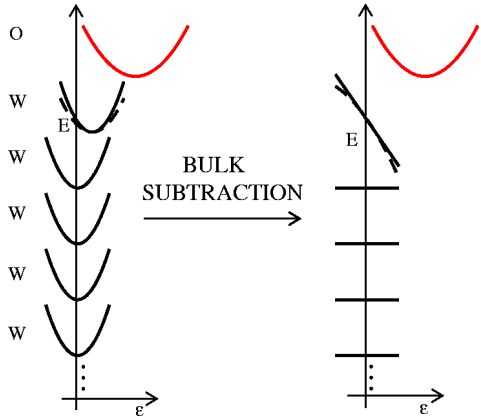}
\caption{ A schematic of energy decomposition in layers before (left) and after the 
subtraction of the bulk energy (right). Red curve represents O surface, and the full 
black line stands for the W layers. The dashed line in the second layer from the top 
represents a case with slightly changed curvature of the W parabola. Dots indicate 
continuation of bulk layers, while  flat lines describe zero energy contribution. 
The horizontal axis represents strain and the vertical axis stands for energy 
(contributions from different layers are drawn with energy shift).    }
\label{fig:parabolas}
\end{center}
\end{figure}

In order to give a heuristic interpretation of Eq.~\ref{eq:stress_layers} and 
\ref{eq:stress_const} in terms of 
energy, in Fig.~\ref{fig:parabolas} we show  the slab energy decomposed in 
layers\cite{note_decomposition} as a function of strain. Horizontal and vertical 
axes represent strain and energy, respectively.  The layer-energy dependence is shown 
shifted in energy, for ease of viewing. The top parabola corresponds to the O surface, 
while those below stand for the subsurface and bulk W layers from top to bottom. For 
simplicity, only the O and the subsurface W layers are shown to be modified with respect 
to bulk W. This decomposition is the energy  analog of Eq.~\ref{eq:stress_layers}. 
For the subsurface W we show two curves, one identical to the bulk curves, and the 
other one with 25~\% smaller curvature.

When we subtract, as is done to determine surface formation energy, the bulk W energy 
from all the W layers, we are left with only two non-zero contributions, from the surface 
and subsurface layers, as shown on the right side of the figure. In the case when the 
subsurface parabola is identical to the bulk one, the remaining contribution after 
subtraction to the surface energy is a line, which will give a constant stress for all 
strains, as written in Eq.~\ref{eq:stress_const}. For a subsurface parabola slightly 
different from the bulk one, the resulting contribution to the surface formation energy 
has a very small curvature, yielding  a slowly-varying stress as a function of strain. 

Finally, adding the energy of the subsurface layer to the surface-layer parabola, we 
obtain a curve comparable to Fig.~\ref{fig:surf_free_energy}, which corresponds to 
the surface stress of the O layer and also contains contribution from the W subsurface 
layer at strain zero.

\acknowledgments
We acknowledge support for this work by the INFM-CNR and CINECA within the 
framework ``Iniziativa  Calcolo per la Fisica della Materia''.


\bibliography{pdo_stripe}

\begin{thebibliography}{44}
\bibitem[{tem()}]{temp_0}
\bibinfo{note}{Disordered systems, for example, can be dealt with using DFT in
  a straighforward manner, but at a price of creating very large supercells and
  various configurations which are to be averaged. This is often impractical or
  impossible due to limited computer resources}.
\bibitem[{not({\natexlab{c}})}]{note_darkfield}
\bibinfo{note}{A {\em dark field} image is obtained with electrons that give
  rise to a particular diffraction spot, which are selected by an aperture
  placed in the back-focal plane of the microscope.}
\bibitem[{not({\natexlab{d}})}]{note_O2}
\bibinfo{note}{$E_{O_2}$ is calculated for an isolated, spin-polarized molecule
  in a large supercell.}
\bibitem[{not({\natexlab{e}})}]{note_biaxial}
\bibinfo{note}{We write all equations for the case of uniaxial stress, which we
  use for our analysis, but the general tensor stress components, $\tau_{i,j}$
  can be used in the same expressions in place of uniaxial stresses. We note
  that our calculations gave us non-diagonal stresses always smaller than 10~\%
  of the value of the diagonal terms.}
\bibitem[{not({\natexlab{f}})}]{note_area}
\bibinfo{note}{In our calculational cell, $A=13.94~\AA^2$ at the equilibrium
  lattice constant.}
\bibitem[{not({\natexlab{g}})}]{note_W_disturbed}
\bibinfo{note}{Our previous DFT results\cite{MenStoBin08} indicate that the W
  atomic positions are only slightly modified upon O adsorption, both laterally
  and along the surface normal.}
\bibitem[{not({\natexlab{h}})}]{note_subsurf}
\bibinfo{note}{In Eq.~\ref{eq:stress_layers} we could have chosen any number of
  subsurface layers to be considered separately. They would all give a
  contribution to the surface stress in Eq.~\ref{eq:stress_const} at the zero
  strain.}
\end{thebibliography}
\end{document}